\documentclass[aps,pra,reprint,amsmath,amssymb,superscriptaddress,longbibliography]{revtex4-2} 
\usepackage{color,graphicx,float,hyperref,stackengine,braket,bm,mathtools,calc,float,eqnarray}
\usepackage[caption=false]{subfig}

\begin{document}
\title{Exploring Nonreciprocal Noise Transfer under Onsager-Casimir Symmetry \\in Synthetic-Field Optomechanics}
\newcommand{\bilkent}{Department of Physics, Bilkent University, 06800 Çankaya, Ankara, Turkey}
\author{Beyza S\"{u}tl\"{u}o\u{g}lu Ege}
 \affiliation{\bilkent}
\author{\c{S}ahin K. {\"{O}}zdemir}%
\affiliation{ 
Department of Electrical and Computer Engineering, Saint Louis University, St. Louis, Missouri 63103, USA
}%
\author{Ceyhun Bulutay}
\email{bulutay@fen.bilkent.edu.tr}
\affiliation{\bilkent}
\date{\today}
\begin{abstract}
An optomechanical system of fundamental importance consists of two intercoupled mechanical resonators, which are radiation-pressure coupled individually to a photonic cavity. This closed-loop and overall lossy configuration possesses two exceptional points (EPs) and offers the realization of synthetic magnetism, controlled by the loop phase. To elucidate the intricate role of loop phase and EPs in this setting, we analyze the noise power spectral density profiles of internal as well as output fluctuations. In the presence of a synthetic magnetic field, the nonreciprocal routing of a signal is well known. Here, we further show that this also applies to nonreciprocal backaction noise flow when the time-reversal symmetry is broken, while the Onsager-Casimir symmetry still holds. To better quantify this phenomenon, we introduce a nonreciprocity measure that contrasts the time-reversed counterparts as a function of loop phase. We observe that nonreciprocal noise flow is enhanced for smaller intermechanical couplings at the expense of lower sensitivity, whereas for sensing purposes, using a higher intermechanical coupling constant is the more viable option. 
\end{abstract}


\maketitle

\section{Introduction}
Open quantum systems described by non-Hermitian Hamiltonians with complex eigenvalue spectrum attract attention due to their aptness for modeling realistic systems in optics \cite{el2018non,el2019dawn,wang2023non,miri2019exceptional,ozdemir2019parity}. When a non-Hermitian system is steered in the parameter space, it can brought to an exceptional point (EP)---a non-Hermitian singularity with coalescing eigenvalues and eigenstates---that are special type of spectral singularities which open a venue for applications and technologies such as enhanced responsivity in optical and electromechanical sensors \cite{hodaei2017enhanced,chen2017exceptional,lu2021exceptional}, realizing high-sensitivity measurements \cite{wu2023chip}, optomechanically induced transparency \cite{PhysRevA.104.033504,jing2015optomechanically, jiao2016nonlinear}, phonon lasing \cite{jing2014pt}, adiabatic transport \cite{hoeller2020non}, photonic quantum interference \cite{klauck2025crossing}, and enhanced cooling and squeezing \cite{sutluouglu2024selective}.  A particularly interesting example of non-Hermitian systems is unidirectional transport \cite{ramezani2010unidirectional,peng2014parity,aleahmad2017integrated}.  Such nonreciprocal and chiral effects have garnered significant attention in optics and photonics  \cite{xu2016topological,reisenbauer2024non,gou2020tunable,choi2017extremely,laha2020exceptional,lau2018fundamental,lee2023chiral,soleymani2022chiral},  atomic and trapped-ion systems \cite{PhysRevA.109.063329, bu2024chiral,zhang2025chirality}, and metamaterials \cite{PhysRevLett.124.257403, PhysRevLett.128.013902,schomerus2020nonreciprocal,PhysRevResearch.3.L022006}.

A complementary, but conceptually rich framework for achieving nonreciprocity invokes the use of artificial gauge fields, especially in optomechanical systems where photons interact with mechanical vibrations, so that these systems become a suitable platform for simulating synthetic gauge fields \cite{schmidt2015optomechanical,walter2016classical,chen2021synthetic, mathew2020synthetic,ehrhardt2023perspective}. Interaction between mechanical and optical modes via radiation pressure allows phase-sensitive coupling, generating complex hopping amplitudes by tuning the phases of the external driving lasers \cite{chen2021synthetic}. These gauge fields in the form of a complex phase factor introduce U(1) gauge
symmetry in photonic and optomechanical systems \cite{aidelsburger2018artificial}. Introducing a synthetic gauge field with a nonzero phase around a plaquette breaks the time-reversal symmetry, leading to nonreciprocity in photonic and optomechanical systems \cite{hafezi2012optomechanically,manipatruni2009optical,xu2020nonreciprocity}. Nonreciprocity enables robust, direction-dependent signal transport, which is essential for integrated photonic circuits, particularly in the implementation of nonreciprocal devices such as isolators, modulators, and circulators \cite{miri2017optical,xu2015optical, metelmann2015nonreciprocal, shen2016experimental,lee2025chiral}. Breaking reciprocity symmetries, inevitably opens additional noise pathways which couple into the measurement process \cite{xu2025enhancing,ye2025routing}. Building on the discussions of $U(1)$ gauge symmetry and artificial gauge fields that are the key mechanisms for nonreciprocity, a fundamental concept is the Onsager-Casimir relation. This is the generalization of the Onsager relation, which states that linear response of a system near equilibrium is symmetric under time-reversal symmetry and microscopic reversibility \cite{onsager1931reciprocal1, onsager1931reciprocal2}. This relation turns into Onsager-Casimir relation when time-reversal symmetry is broken \cite{brechet2022onsager}.

A significant obstacle in high-precision optical measurements is handling the introduced noise in the system. Optical systems’ performance of sensitivity contend two intertwined fundamental sources of uncertainty: One of them is induced with the fluctuations in the probe laser (imprecision or the so-called shot noise), and the other is the backaction noise introduced on the light while probing a system (e.g., radiation-pressure noise in optomechanical systems). A primary challenge in quantum sensing is to identify the methods to circumvent both shot noise and measurement backaction noise. Various techniques have been proposed to overcome those limitations, including the use of squeezed light, \cite{lee2020squeezed,PhysRevX.3.031012} backaction-evading measurements schemes \cite{shomroni2019optical,hertzberg2010back}, and the coherent quantum noise cancellation \cite{tsang2010coherent, wimmer2014coherent, Bariani_2015}. While all these methods ultimately aim to enhance measurement sensitivity by alleviating noise, they operate through distinct physical mechanisms. A theoretical framework for ponderomotive squeezing of light through the radiation-pressure interaction between a laser field and a cryogenically cooled membrane achieved a 32\% sensitivity below the shot-noise level \cite{PhysRevX.3.031012}. Another method is directly injecting a coherent superposition of coherent light and squeezed vacuum field \cite{lee2020squeezed} or frequency-dependent squeezed vacuum into a quantum sensing system \cite{subhash2022enhancing}. While these squeezed-light utilization methods reduce the imprecision noise by adjusting the input noise, alternative methods focus on backaction-evading measurements by directly suppressing the radiation-pressure noise. These methods utilize the fact that radiation pressure predominantly couples to a certain mechanical quadrature, allowing backaction evasion techniques to enable continuous measurement without disturbing the conjugate variable, and up to 0.67~dB (14\%) reduction of total measurement noise can be achieved \cite{shomroni2019optical}. Additionally, backaction noise across all frequencies can be eliminated via quantum coherent-noise cancellation by identifying interference pathways between quantum noises \cite{tsang2010coherent}, adding an auxiliary optical cavity mimicking a negative-mass oscillator \cite{wimmer2014coherent}, or using an ensemble of ultracold atoms, which provides a negative-mass dynamics \cite{Bariani_2015}.
Beyond these efforts for minimizing measurement noise, recent research focused on asymmetry in signal flow, particularly nonreciprocity, which is necessary for amplifiers and isolators. Interestingly, thermal-noise cancellation can be used for achieving optomechanically induced nonreciprocity when a whispering-gallery-mode cavity is coupled to mechanical resonators. The system can be designed so that thermal-noise pathways from each mechanical resonator interfere destructively and cancel thermal noise in the system, culminating in asymmetric signal transmission and low thermal noise without needing precooling mechanisms \cite{tang2023thermal}. 

In this article, we study theoretically a closed-loop optomechanical system consisting of a photonic cavity and intercoupled identical lossy mechanical resonators. This configuration, constituting a synthetic magnetism, enables a controllable nonreciprocal noise transfer mediated by the closed-loop phase. This is achieved when time-reversal symmetry is broken while the Onsager-Casimir relation is still valid \cite{brechet2022onsager}. First, by exploring the intracavity and cavity-output power spectral density (PSD), we demonstrate noise cancellation at the upper and lower bands depending on the loop phase. Next, we show that the internal noise PSD of mechanical resonators can be independently controlled, which paves the way for nonreciprocity. The most conspicuous outcome is the nonreciprocal backaction noise transfer between the mechanical resonators, and this flow is enhanced for smaller mechanical coupling constants. This optomechanical scheme can also be realized as a sensor with increased sensitivity beyond the first EP.
\begin{figure}[H]
  \centering
  \includegraphics[width=\linewidth]{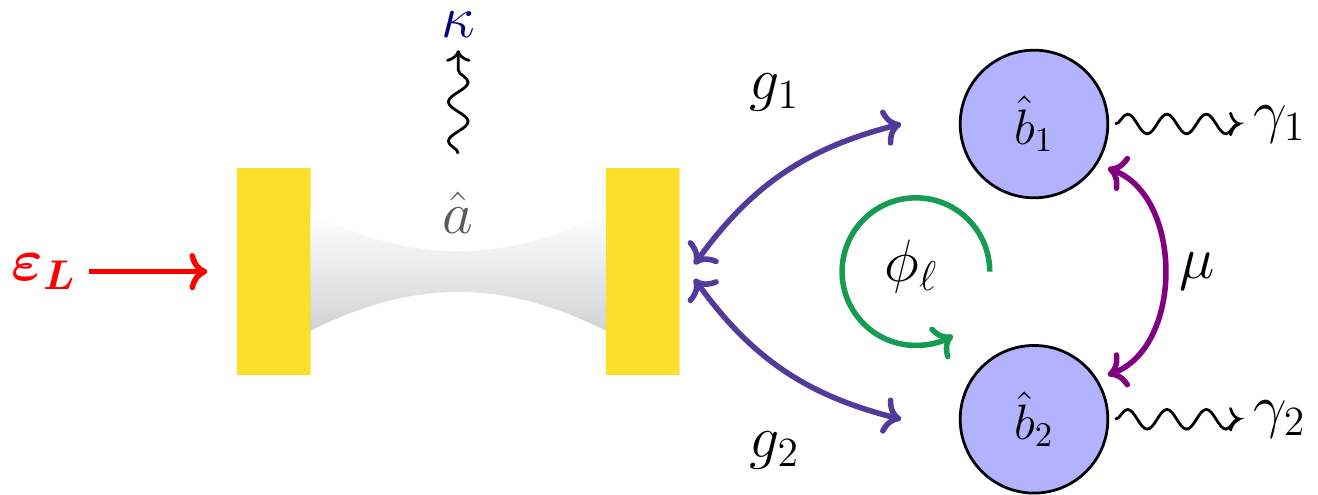}
      \caption{Non-Hermitian closed-loop optomechanical system composed of a photonic cavity and two lossy mechanical resonators with the resonance frequencies $\omega_{c}$ and $\omega_m$, respectively. The photonic cavity is pumped via a coherent laser with amplitude $\varepsilon_{L}$. Loss rates are indicated with wavy arrows, and coupling rates with solid arrows.  }
    \label{fig1}
\end{figure}
\section{Theory}
We consider a closed-loop optomechanical system consisting of a photonic cavity coupled to two lossy mechanical resonators via radiation-pressure coupling constants $g_1$ and $g_2$ as shown in Fig.~\ref{fig1}. Loss rates of first and second mechanical resonators are $\gamma_1>0$ and $\gamma_2 >0$, respectively. Both mechanical resonators share an equal resonance angular frequency $\omega_m$, and are intercoupled via mechanical coupling constant $\mu$. The photonic cavity is driven with a coherent pump laser with carrier frequency $\omega_L$ and amplitude $\varepsilon_L = \sqrt{\frac{2P_L}{\hbar \omega_L}}$, which is associated with the corresponding laser power $P_L$. 
\subsection{ Onsager-Casimir relation}
Onsager relations are established in the context of statistical physics, and in general, they assert that the linear response of a system is symmetric with respect to input and output. Specifically, when time-reversal symmetry and microscopic reversibility holds, Onsager symmetry guarantees that the linear response matrix entries, $L_{ij}$ have symmetric cross-couplings between paired fluxes and forces, i.e., $L_{ij}=L_{ji}$ \cite{onsager1931reciprocal1,onsager1931reciprocal2}. Onsager reciprocity relations are widely applicable in classical thermodynamics, and in electromagnetic systems are known under the name Lorentz reciprocity \cite{caloz2018electromagnetic}. However, when time-reversal symmetry is broken, say under magnetic fluxes, synthetic fields, nonlinearity, or gain/loss engineering, these symmetry relations are generalized into Onsager-Casimir relations. They account for time-reversal parity of the associated quantities ($\varepsilon_{i}=+1$ for even, and $\varepsilon_{i}=-1$ for odd parity), and impose a modified symmetry $L_{ij}(B) = \varepsilon_i \varepsilon_j L_{ji}(-B)$, which mandates that the system’s dynamics remain microscopically reversible under magnetic field $(B)$ reversal \cite{brechet2022onsager}. This generalization forms the theoretical foundation for understanding and designing non-Hermitian and nonreciprocal photonic devices \cite{tsang2024quantum, vicencio2024non}. For our optomechanical system, where the closed-loop coupling phase with $U(1)$ symmetry introduces synthetic magnetism, the generalized Onsager-Casimir relations hold. In other words, the response is invariant if the mechanical resonators are interchanged ($1\leftrightarrow 2$) together with negating the loop phase, i.e., $\phi_\ell\rightarrow -\phi_\ell$.
\subsection{The optomechanical Hamiltonian}
The Hamiltonian in the rotating frame of the pump laser at frequency $\omega_{L}$ ($\hbar=1$) is
\begin{eqnarray}
\nonumber
\hat{H} & = &  \Delta \hat{a}^\dagger \hat{a} + \omega_m (\hat{b}_1^\dagger \hat{b}_1 + \hat{b}_2^\dagger \hat{b}_2)- (\mu\hat{b}_1^\dagger \hat{b}_2 + \mu^*\hat{b}_2^\dagger \hat{b}_1)  \\
 \nonumber
& &-\hat{a}^\dagger \hat{a}( g_1\hat{b}_1^\dagger + g_1^*\hat{b}_1)-\hat{a}^\dagger \hat{a}(g_2 \hat{b}_2^\dagger + g_2^* \hat{b}_2) \\
\nonumber
& &+ i \sqrt{\eta \kappa} \left[\varepsilon_L \hat{a}^\dagger-\varepsilon^*_L \hat{a}\right].
\label{Hamiltonian}
 \end{eqnarray} 
Here, $\Delta= \omega_{c}-\omega_L$ is the detuning between input laser and the cavity, $\hat{a}~(\hat{a}^\dagger)$, and $\hat{b}_1~(\hat{b}_1^\dagger)$, and $\hat{b}_2~(\hat{b}_2^\dagger)$ are the annihilation (creation) operators of the optical and mechanical modes, respectively, $\kappa$ is the cavity decay rate, and $\eta$ is the escape efficiency. 
\subsection{Quantum Langevin equations}
This Hamiltonian formalism is suitable for closed systems, whereas we would like to account for the losses and input noises.
Thus, from the above Hamiltonian, we obtain the following quantum Langevin equations, which govern the time evolution of the modes
\begin{eqnarray}
\nonumber
\frac{d\hat{a} }{dt}&=& -i\Delta \hat{a} + i \hat{a}(g_1 \hat{b}_1^\dagger + g_1^* \hat{b}_1) + i\hat{a} (g_2 \hat{b}_2^\dagger + g_2^* \hat{b}_2)\\
\nonumber
& &+ \eta\kappa \varepsilon_L -\frac{\kappa}{2}\hat{a} +\sqrt{\kappa}\hat{a}_{\mathrm{in}}(t),\\
\nonumber
\frac{d\hat{b}_1}{dt} &=& -i\omega_m \hat{b}_1 +i\mu\hat{b}_2 +i g_1 \hat{a}^\dagger \hat{a} -\frac{\gamma_1}{2} \hat{b}_1 +\sqrt{\gamma}_1 \hat{b}_{1,\mathrm{in}}(t),\\
\nonumber
\frac{d\hat{b}_2}{dt} &=& -i\omega_m \hat{b}_2 +i\mu^*\hat{b}_1 +i g_2 \hat{a}^\dagger \hat{a} -\frac{\gamma_2}{2} \hat{b}_2 +\sqrt{\gamma}_2 \hat{b}_{2,\mathrm{in}}(t),
\end{eqnarray}
 where $\hat{a}_{\mathrm{in}}(t)$, $\hat{b}_{1,\mathrm{in}}(t)$, and $\hat{b}_{2,\mathrm{in}}(t)$ are zero-mean cavity input noise and mechanical thermal noise operators, respectively \cite{gardiner2004quantum}. Photonic and mechanical Markovian input noise operators satisfy the following non-zero correlations
 \begin{eqnarray}
\langle \hat{a}_{\mathrm{in}}^\dagger(t) \hat{a}_{\mathrm{in}}(t') \rangle &=& n_a \delta(t-t'),\\
\nonumber
\langle \hat{a}_{\mathrm{in}}(t) \hat{a}_{\mathrm{in}}^\dagger(t') \rangle &=& (n_a +1) \delta(t-t'),\\
\nonumber
\langle \hat{b}_{j,\mathrm{in}}^\dagger(t) \hat{b}_{j,\mathrm{in}}(t') \rangle &=& n_{b_j} \delta(t-t'),\\
\nonumber
\langle \hat{b}_{j,\mathrm{in}}(t) \hat{b}_{j,\mathrm{in}}^\dagger(t') \rangle &=& (n_{b_j}+1) \delta(t-t'),
\end{eqnarray}
for $j=1, 2$, and where $n_a$ and $n_{b_j}$ are the mean thermal occupancies of the optical cavity and respective thermal reservoirs \cite{clerk2010introduction}. Since our main focus is on fluctuations around the steady-state mean values, subsequently we linearize quantum Langevin equations by expanding each mode as a sum of the classical mean value and quantum fluctuation, $\hat{\aleph}(t)\rightarrow \langle\hat{\aleph}(t)\rangle + \delta\hat{\aleph}(t)$, where $\hat{\aleph}=\hat{a}, \hat{b}_1, \hat{b}_2$. The steady-state mean values are given by
\begin{eqnarray}
\nonumber
\langle \hat{a} \rangle &=& \frac{\sqrt{\eta\kappa}\varepsilon_L}{\frac{\kappa}{2}+i\Delta_a},\\
\nonumber
\langle \hat{b}_1 \rangle &=& \frac{i\mu \langle \hat{b}_2\rangle + ig_1|\langle \hat{a} \rangle|^2}{\frac{\gamma_1}{2}+i\omega_m},\\
\nonumber
\langle \hat{b}_2 \rangle &=& \frac{i\mu^* \langle \hat{b}_1\rangle + ig_2|\langle \hat{a} \rangle|^2}{\frac{\gamma_2}{2}+i\omega_m},\\
\nonumber
\end{eqnarray}
where mechanical motion mediated detuning is defined as  $\Delta_a \coloneq \Delta -2 \operatorname{Re}\left(g_1\langle \hat{b}_1 \rangle^* + g_2^*\langle \hat{b}_2 \rangle\right).$
The quantum fluctuations denoted by $\delta\hat{\aleph}(t)$ around the steady-state mean values are 
\begin{eqnarray}
\label{qfcav}
\nonumber
\frac{d\delta\hat{a}}{dt}&=& i(-\Delta_a+i\kappa/2)\delta\hat{a}+\sqrt{\kappa}\hat{a}_{\mathrm{in}}(t) \\
\nonumber
& & +i\langle \hat{a} \rangle g_1^* \delta \hat{b}_1 +i\langle \hat{a}\rangle g_1 \delta \hat{b}_1^\dagger +i \langle \hat{a} \rangle g_2^* \delta \hat{b}_2 +i\langle \hat{a} \rangle g_2 \delta \hat{ b}_2^\dagger,\\
\nonumber
\label{qfMR1}
\frac{d\delta \hat{b}_1}{dt} &=& i(-\omega_m +i\gamma_1/2)\delta \hat{b}_1 +i\mu\delta \hat{b}_2 + ig_1\langle \hat{a} \rangle^* \delta \hat{a} \\
\nonumber
& & + ig_1\langle \hat{a} \rangle \delta \hat{a}^\dagger + \sqrt{\gamma_1} \hat{b}_{1,\mathrm{in}}(t),\\
\label{qfMR2}
\nonumber
\frac{d\delta \hat{b}_2}{dt} &=& i(-\omega_m +i\gamma_2/2)\delta \hat{b}_2 +i\mu^*\delta \hat{b}_1 + ig_2\langle \hat{a} \rangle^* \delta \hat{a} \\
\nonumber
& & + ig_2\langle \hat{a} \rangle \delta \hat{a}^\dagger + \sqrt{\gamma_2} \hat{b}_{2,\mathrm{in}}(t).
\nonumber
\label{quantum fluctuation}
\end{eqnarray}
Next, we introduce quadrature operators of position and momentum for cavity and mechanical modes in terms of fluctuation operators as
\begin{eqnarray}
\nonumber
\delta \hat{X}_{\aleph=a,b_1,b_2} &=&  \frac{\delta \hat{\aleph} +\delta\hat{\aleph}^\dagger }{\sqrt{2}}, \\
\nonumber
\delta \hat{Y}_{\aleph=a,b_1,b_2} &=&  \frac{\delta \hat{\aleph} -\delta\hat{\aleph}^\dagger }{i \sqrt{2}},
\nonumber
\end{eqnarray}
where the corresponding quadrature input noise operators are
\begin{eqnarray}
\nonumber
\hat{X}^{\mathrm{in}}_{\aleph=a,b_1,b_2} &=&  \frac{\hat{\aleph}_{\mathrm{in}} +\hat{\aleph}_{\mathrm{in}} ^\dagger }{\sqrt{2}},\\
\nonumber
\hat{Y}^{\mathrm{in}}_{\aleph=a,b_1,b_2} &=&  \frac{\hat{\aleph}_{\mathrm{in}} -\hat{\aleph}_{\mathrm{in}} ^\dagger }{i \sqrt{2}}.
\end{eqnarray}
These position-momentum quadrature fluctuation operators can be cast into a matrix equation as
\begin{equation}
\mathbf{\dot{\hat{R}}}(t) = \mathbf{M} \mathbf{\hat{R}}(t) + \mathbf{\hat{N}}(t),
\label{quadraturefluc eqn}
\end{equation}
where over dot represents time derivative, and $\mathbf{\hat{R}}(t) =[\delta \hat{X}_{a},\delta\hat{Y}_{a},\delta\hat{X}_{b_1},\delta\hat{Y}_{b_1},\delta\hat{X}_{b_2},\delta\hat{Y}_{b_2}]^T$ stands for quadrature fluctuation operators, and $\mathbf{M}$ is a 6$\times$6 time-independent drift matrix, which consists of quantum fluctuation coefficients
\begin{widetext}
\begin{equation}
\nonumber
\resizebox{\textwidth}{!}{$
\mathbf{M} = \begin{pmatrix}
-\frac{\kappa}{2} & \Xi & -2\operatorname{Re}(g_1)\operatorname{Im}(\langle\hat{a}\rangle) & -2\operatorname{Im}(g_1)\operatorname{Im}(\langle\hat{a}\rangle) & -2\operatorname{Re}(g_2)\operatorname{Im}(\langle\hat{a}\rangle) & -2\operatorname{Im}(g_2)\operatorname{Im}(\langle\hat{a}\rangle) \\
-\Xi & -\frac{\kappa}{2} & 2\operatorname{Re}(g_1)\operatorname{Re}(\langle\hat{a}\rangle) & 2\operatorname{Im}(g_1)\operatorname{Re}(\langle\hat{a}\rangle) & 2\operatorname{Re}(g_2)\operatorname{Re}(\langle\hat{a}\rangle) & 2\operatorname{Im}(g_2)\operatorname{Re}(\langle\hat{a}\rangle) \\
-2\operatorname{Im}(g_1)\operatorname{Re}(\langle\hat{a}\rangle) & -2\operatorname{Im}(g_1)\operatorname{Im}(\langle\hat{a}\rangle) & -\frac{\gamma_1}{2} & \omega_m & -\operatorname{Im}(\mu) & -\operatorname{Re}(\mu) \\
2\operatorname{Re}(g_1)\operatorname{Re}(\langle\hat{a}\rangle) & 2\operatorname{Re}(g_1)\operatorname{Im}(\langle\hat{a}\rangle) & -\omega_m & -\frac{\gamma_1}{2} & \operatorname{Re}(\mu) & -\operatorname{Im}(\mu) \\
-2\operatorname{Im}(g_2)\operatorname{Re}(\langle\hat{a}\rangle) & -2\operatorname{Im}(g_2)\operatorname{Im}(\langle\hat{a}\rangle) & \operatorname{Im}(\mu) & -\operatorname{Re}(\mu) & -\frac{\gamma_2}{2} & \omega_m \\
2\operatorname{Re}(g_2)\operatorname{Re}(\langle\hat{a}\rangle) & 2\operatorname{Re}(g_2)\operatorname{Im}(\langle\hat{a}\rangle) & \operatorname{Re}(\mu) & \operatorname{Im}(\mu) & -\omega_m & -\frac{\gamma_2}{2}
\end{pmatrix}
$},
\end{equation}
\end{widetext}
where for notational convenience we define $\Xi \coloneq \Delta - 2(\operatorname{Re}(g_1)\operatorname{Re}(\langle\hat{b}_1\rangle) + \operatorname{Re}(g_2)\operatorname{Re}(\langle\hat{b}_2\rangle)) - 2(\operatorname{Im}(g_1)\operatorname{Im}(\langle\hat{b}_1\rangle) + \operatorname{Im}(g_2)\operatorname{Im}(\langle\hat{b}_2\rangle))$. Noise vector $\mathbf{\hat{N}}(t)$ is defined as 
\begin{equation}
\begin{aligned}
\mathbf{\hat{N}}(t)
= \big[&
\sqrt{\kappa}\hat{X}^{\mathrm{in}}_{a},
\sqrt{\kappa}\hat{Y}^{\mathrm{in}}_{a},
\sqrt{\gamma_1}\hat{X}^{\mathrm{in}}_{b_1},\sqrt{\gamma_1}\hat{Y}^{\mathrm{in}}_{b_1},
\sqrt{\gamma_2}\hat{X}^{\mathrm{in}}_{b_2}, \\
&
\nonumber
\sqrt{\gamma_2}\hat{Y}^{\mathrm{in}}_{b_2}
\big]^T,
\end{aligned}
\end{equation}

which is the product of the input noise vector, $\mathbf{\hat{R}}^{\mathrm{in}}(t)$, 
\begin{equation}
\nonumber
\mathbf{\hat{R}}^{in}(t)=\left[\hat{X}^{\mathrm{in}}_{a},\hat{Y}^{\mathrm{in}}_{a},\hat{X}^{\mathrm{in}}_{b_1},\hat{Y}^{\mathrm{in}}_{b_1},\hat{X}^{in}_{b_2},\hat{Y}^{\mathrm{\mathrm{in}}}_{b_2}\right]^T,
\end{equation}
and a diagonal coefficient matrix $\mathbf{D}$ as
\begin{equation}
\mathbf{D} = \mathrm{diag}\left[ \sqrt{\kappa}, \sqrt{\kappa}, \sqrt{\gamma_1}, \sqrt{\gamma_1}, \sqrt{\gamma_2}, \sqrt{\gamma_2}\right],  
\end{equation}
whose diagonal entries represent the individual loss rates associated with each resonator.
\subsection{Internal fluctuations}
We first analyze the internal fluctuations within the resonators and their phase dependence before discussing the output spectrum fluctuations. Transforming Eq.~(\ref{quadraturefluc eqn}) into frequency domain we obtain
\begin{eqnarray}
\label{frequencydomain}
\left[i\omega\mathbf{I} + \mathbf{M}\right] \mathbf{\hat{R}}(\omega)&=&-\mathbf{\hat{N}}(\omega),\, \\
\nonumber
\mathbf{\hat{R}}(\omega)&=&-\mathbf{T}(\omega)\,\mathbf{\hat{N}}(\omega),\, 
\nonumber
\label{fluctuations in frequency}
\end{eqnarray}
where $\mathbf{T}(\omega) \coloneq\left[i\omega\mathbf{I} + \mathbf{M}\right]^{-1}$. Applying a similarity transformation that directly diagonalizes the drift matrix $\mathbf{M}$, the explicit frequency dependence of $\mathbf{T}(\omega)$ matrix is found as
\begin{eqnarray}
\label{T_matrix}
T_{ij}(\omega)=\sum_{k}\frac{U_{ik}U^{-1}_{kj}}{i\omega+\lambda_k},
\end{eqnarray}
where $\mathbf{U}$ is composed of eigenvectors of $\mathbf{M}$. 
Our main aim here is to find the PSD matrix for internal fluctuations given by 
\begin{eqnarray}
\nonumber
\mathbf{S}(\omega) = \langle \hat{\mathbf{R}}(-\omega) \hat{\mathbf{R}}^T(\omega) \rangle. 
\end{eqnarray}
Negative frequency component in PSD definition is obtained by Hermitian conjugate relation of operators, $\hat{\mathcal{O}}^\dagger(\omega)=\hat{\mathcal{O}}(-\omega)$ \cite{bowen2015quantum}. Diagonal elements of $\mathbf{S}(\omega)$ correspond to internal fluctuations of cavity and mechanical modes, respectively. Combining Eqs.~(\ref{frequencydomain}) and (\ref{T_matrix}),
\begin{eqnarray}
\nonumber
\mathbf{S}(\omega) &=& 2\pi \mathbf{T}(-\omega) \mathbf{C}\mathbf{T}^T(\omega), \\
S_{ij}(\omega) &=& \sum_{\ell,q} \sum_{k',k}2\pi \frac{U_{ik}U_{k\ell}^{-1}C_{\ell q}U_{k'q}^{-1}U_{jk'}}{(-i\omega +\lambda_k)(i\omega+\lambda_{k'})},
\nonumber
\end{eqnarray}
in which $\mathbf{C}$ is two-time noise correlation matrix with elements $\langle \mathbf{\hat{N}}(\omega)\mathbf{\hat{N}}^T(\omega')\rangle=2\pi\mathbf{C}\delta(\omega+\omega')$ given as, 
\vspace{1em} 
\begin{widetext}
\begin{equation}
\mathbf{C}=
\begin{pmatrix}
\frac{\kappa}{2}(2n_a+1) & \frac{-\kappa}{2i}&0&0&0&0 \\
\frac{\kappa}{2i} & \frac{\kappa}{2}(2n_a+1)  &0&0&0&0 \\
0&0& \frac{\gamma_1}{2}(2n_m+1)&\frac{-\gamma_1}{2i}&0&0\\
0&0&  \frac{\gamma_1}{2i}&\frac{\gamma_1}{2}(2n_m+1)&0&0\\
0&0&0&0&\frac{\gamma_2}{2}(2n_m+1)&\frac{-\gamma_2}{2i} \\
0&0&0&0&\frac{\gamma_2}{2i}&\frac{\gamma_2}{2}(2n_m+1) \\
\end{pmatrix}.
\end{equation}
\end{widetext}
\subsection{Individual contributions to internal and output fluctuations}
To keep track of the shot and backaction noise
contributions within internal PSD, we begin by expressing the quadrature fluctuations as linear combinations of the input noise operators, weighted by the corresponding susceptibilities in the optomechanical plaquette, as in 
\begin{eqnarray}
\nonumber
\hat{R}_{i} &=& \chi_{R_i Y_a}\hat{Y}_a^{in} + \chi_{R_i X_a}\hat{X}_a^{in} + \chi_{R_i X_{b_1}}\hat{X}_{b_1}^{in} + \chi_{R_i Y_{b_1}}\hat{Y}^{in}_{b_1}  \\
\nonumber
& & + \chi_{R_i X_{b_2}}\hat{X}_{b_2}^{in} +\chi_{R_i Y_{b_2}}\hat{Y}_{b_2}^{in},\\
\nonumber
\end{eqnarray}    
where $R_i\in \{X_a, Y_a, X_{b_1}, Y_{b_1}, X_{b_2}, Y_{b_2}\}$. PSD is defined as ${\mathbf{S}(\omega) = \langle \hat{\mathbf{R}}(-\omega) \hat{\mathbf{R}}^T(\omega) \rangle}$, so the subsequent task is to determine the relevant susceptibilities. To do this, we write Eq.~(\ref{frequencydomain}) as
\begin{eqnarray}
\nonumber
\mathbf{\hat{R}}(\omega) = \mathbf{T^\prime(\omega)}\mathbf{\hat{R}}^{in},
\end{eqnarray}
where the susceptibility matrix is defined as ${\mathbf{T^\prime}(\omega) = \mathbf{T}(\omega)\mathbf{D}}$. 
\subsection{Nonreciprocity measure}
To quantify the noise flow among the resonators, we define a loop phase-dependent nonreciprocity measure, $I_\Delta(\phi_{\ell})$, in terms of integrated spectral densities as
\begin{align}
F_{j \rightarrow i}(\phi_{\ell},\omega)
&\coloneq
\bigl|T^\prime_{R_{i}R_{j}}(\phi_{\ell},\omega)\bigr|^2
S^{\mathrm{in}}_{R_j R_j}(\phi_{\ell},\omega), \nonumber\\
F_{i \rightarrow j}(\phi_{\ell},\omega)
&=
F_{j \rightarrow i}(\phi_{\ell}+\pi,\omega), \nonumber\\
I_\Delta(\phi_{\ell})
&\coloneq
\left|
\int_{-\infty}^{\infty}
\!\!\left[
F_{j \rightarrow i}(\phi_{\ell},\omega)
-
F_{i \rightarrow j}(\phi_{\ell},\omega)
\right]
\,d\omega
\right|,
\label{non-reciprocity measure}
\end{align}
where $F_{j \rightarrow i}(\phi_{\ell},\omega)$ and $F_{j \rightarrow i}(\phi_{\ell}+\pi,\omega)$ are the time-reversed counterparts of each other and represent the noise transfer from $R_j$ to $R_i$ quadrature. Input noise PSD matrix is defined as $\mathbf{S}^{\mathrm{in}}(\omega) =  \langle\mathbf{\hat{R}}^{\mathrm{in}}(-\omega) \mathbf{\hat{R}}^{\mathrm{in}^T}(\omega) \rangle$. This way we characterize nonreciprocity in physical terms as an imbalance in the noise flow between the quadratures $R_i$ and $R_j$. Notably, for the $\phi_{\ell}=0$ or $\pi$ values, time-reversal symmetry is restored, as will be demonstrated in the results to follow. 
\subsection{Output power spectral density}
For sensing applications, it is necessary to determine the output PSD, which relies on the input-output relationship. To explore the sensitivity, a homodyne measurement in optical mode with a homodyne angle $\theta$ is performed, where a rotated quadrature can be measured depending on the homodyne angle. Namely, if $\theta=0^\circ$, the position quadrature is measured, whereas if $\theta=90^\circ$, the measurement corresponds to the momentum quadrature. By choosing angles other than $0^\circ$ or $90^\circ$, it becomes possible to measure a rotated quadrature \cite{sudhir2017quantum}. Output quadratures of the optical cavity and the mechanical resonators can be expressed in terms of the input noises and corresponding susceptibilities, also called transfer functions \cite{beckey2023quantum}:
\begin{eqnarray}
\nonumber
\hat{R}_{i}^{\mathrm{out}} &=& \chi_{R_i Y_a}\hat{Y}_a^{\mathrm{in}} + \chi_{R_i X_a}\hat{X}_a^{\mathrm{in}} + \chi_{R_i X_{b_1}}\hat{X}_{b_1}^{\mathrm{in}} + \chi_{R_i Y_{b_1}}\hat{Y}^{\mathrm{in}}_{b_1} \\
\nonumber
& & + \chi_{R_i X_{b_2}}\hat{X}_{b_2}^{\mathrm{in}} +\chi_{R_i Y_{b_2}}\hat{Y}_{b_2}^{\mathrm{in}}.\\
\nonumber
\end{eqnarray}    
The output PSD matrix is defined as
\begin{eqnarray}
\nonumber
\mathbf{S}^{\mathrm{out}}(\omega) =  \langle\mathbf{\hat{R}}^{\mathrm{out}}(-\omega) \mathbf{\hat{R}}^{\mathrm{out}^T}(\omega) \rangle.
\end{eqnarray}
To proceed, the output quadrature operators must be expressed in terms of the input noises as
\begin{eqnarray}
\left[i\omega\mathbf{I} + \mathbf{M}\right] \mathbf{\hat{R}}(\omega)&=&-\mathbf{D}\mathbf{\hat{R}}^{\mathrm{in}}(\omega).
\label{differential eqn in frequency}
\end{eqnarray}
For that purpose, we use a slightly modified version of Eq.~(\ref{frequencydomain}) where the coefficients contained in $\mathbf{\hat{N}}(\omega)$ are transferred into a separate matrix, for the sake of simplicity in calculations. Afterwards, input-output relationship is applied to Eq.~(\ref{differential eqn in frequency}) \cite{clerk2010introduction},
\begin{eqnarray}
\nonumber
\mathbf{\hat{R}}^{\mathrm{out}}(\omega) &=& \mathbf{\hat{R}}^{\mathrm{in}}(\omega)-\mathbf{D}\mathbf{\hat{R}}(\omega),\\
\nonumber
\mathbf{\hat{R}}^{\mathrm{out}}(\omega) &=& \mathbf{P}(\omega)\mathbf{\hat{R}}^{\mathrm{in}},
\end{eqnarray}
where $\mathbf{P}(\omega)=(\mathbf{\hat{I}} + \mathbf{D}\mathbf{T}(\omega)\mathbf{D})$ is the susceptibility matrix. 
\section{The parameter set and its experimental relevance}
\label{sec3}
In this section, we discuss the experimental aspects of our theoretical work before presenting our parameter set. In practical applications, intercoupling two mechanical modes and thereby realizing synthetic magnetism through a loop phase is an experimental challenge \cite{fang2017generalized,del2022non,xu2019nonreciprocal}. In our study, individual coupling phases can be gauged out so that only the global loop phase remains \cite{PhysRevA.104.033504}. Hence, this overall loop phase can be realized by embodying it in the intermechanical coupling constant. For instance, coupling can be conveyed through the optical radiation field whose carrier phase is locked to the pump laser with a controllable phase shift via integrated electro-optic and thermo-optic modulators \cite{gil2017light,zhang2012synchronization}.

The parameters used in this article are adopted from a seminal experimental work, which substantiates the realizability of our optomechanical system \cite{safavi2011electromagnetically}. However, we should note that this study contained a single mechanical resonator, thus no intermechanical coupling constant, unlike our work.
To reduce the parameter space, we consider identical mechanical resonators with the same losses and resonance frequencies. For this purpose, we set, $g_{1,2}\in\mathbb{R}$ with $g_1/(2\pi)=g_2/(2\pi)=800$~kHz, $\omega_m/(2\pi)=3.75$~GHz, $\kappa/(2\pi)=900$~MHz, equal amount of mechanical losses $\gamma_1=\gamma_2= 5 \times 10^{-4}\,\omega_m$, optical cavity escape efficiency $\eta=0.5$, laser power $P$ = 0.125~mW with a wavelength of 1550~nm, and detuning $\Delta = \omega_m$ (red-detuned pumping). This corresponds to the resolved-sideband regime $\kappa<\omega_m$ and weak-coupling limit $g_{1,2}\ll \kappa$. This parameter set possesses two EPs, $|\mu_{\mathrm{EP},1}| \simeq 31.06\,(\gamma_1+ \gamma_2)$  and  $|\mu_{\mathrm{EP},2}| \simeq 41.6\,(\gamma_1+ \gamma_2)$, where both eigenvalues and the eigenvectors coalesce in the parameter space. Numerical values for the EPs are found from the root loci of the eigenvalues, $(z\left(|\mu|,\phi_{\ell}\right)=\alpha+i\omega,~\omega>0)$ of the drift matrix $\mathbf{M}$ ~\cite{sutluouglu2024selective}. Unless stated otherwise, we choose the intermechanical coupling constant value at the second EP, $\mu_{\mathrm{EP},2}$. Therefore, the optomechanical system operates at the strong coupling regime, i.e., $\mu \gg \gamma_{1,2}$  \cite{okamoto2013coherent,luo2018strong,deng2016strongly,mathew2016dynamical}. We assume the optical cavity mode is in the vacuum state, while both mechanical reservoirs have an equal mean number of phonons, $n_{b_1} = n_{b_2} = 100$; this corresponds to an ambient temperature of $T = 18.1~$K.

\section{Results}
\begin{figure}[H]
  \centering
  \includegraphics[width=\linewidth]{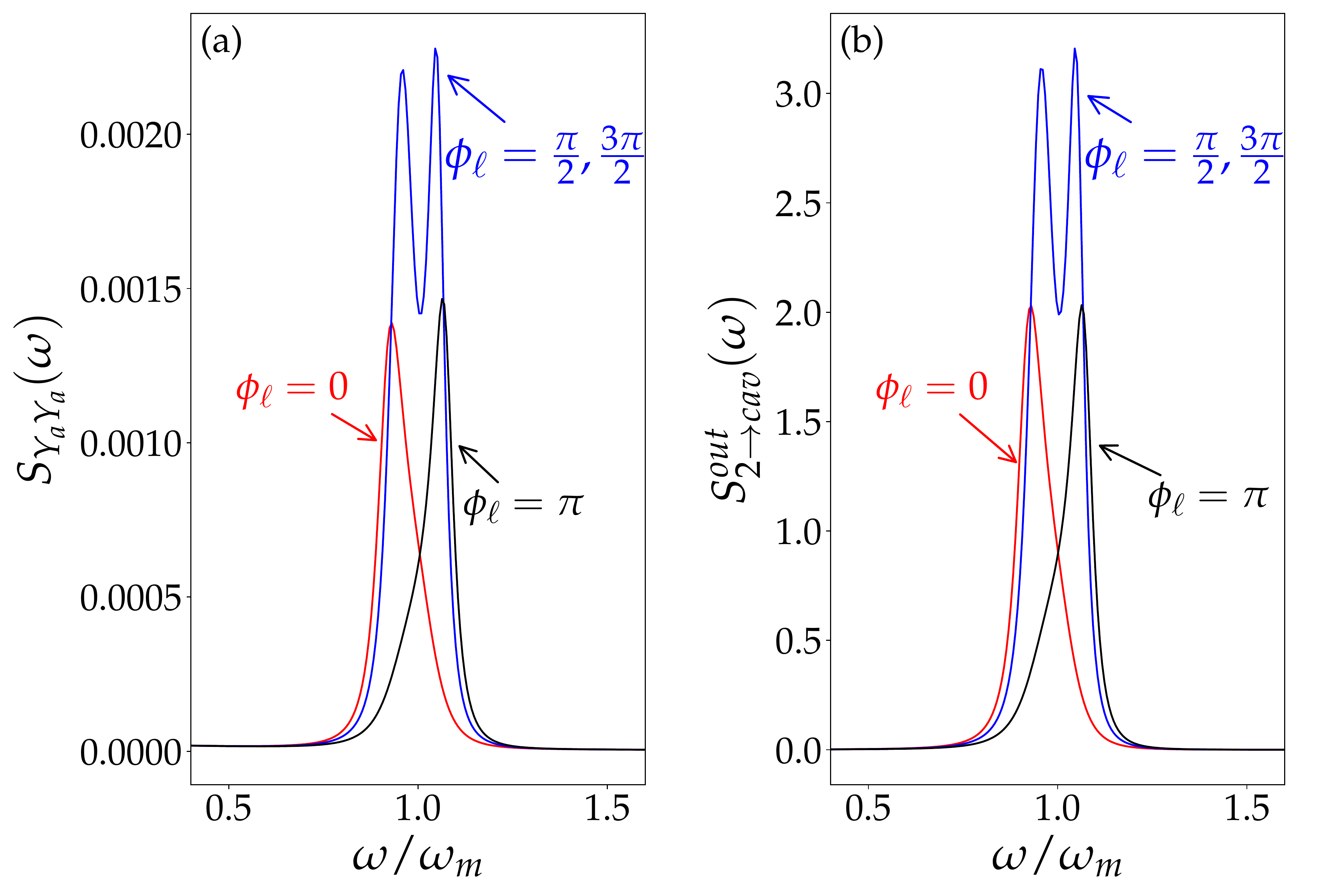}
      \caption{(a) Optical intracavity PSD, (b) noise contribution of the second mechanical resonator to the cavity-output noise spectrum for loop phases, $\phi_{\ell} = 0,~\pi/2,~\pi$, and $3\pi/2$. }
    \label{fig2}
\end{figure}

\begin{figure}[H]
  \centering
  \includegraphics[width=\linewidth]{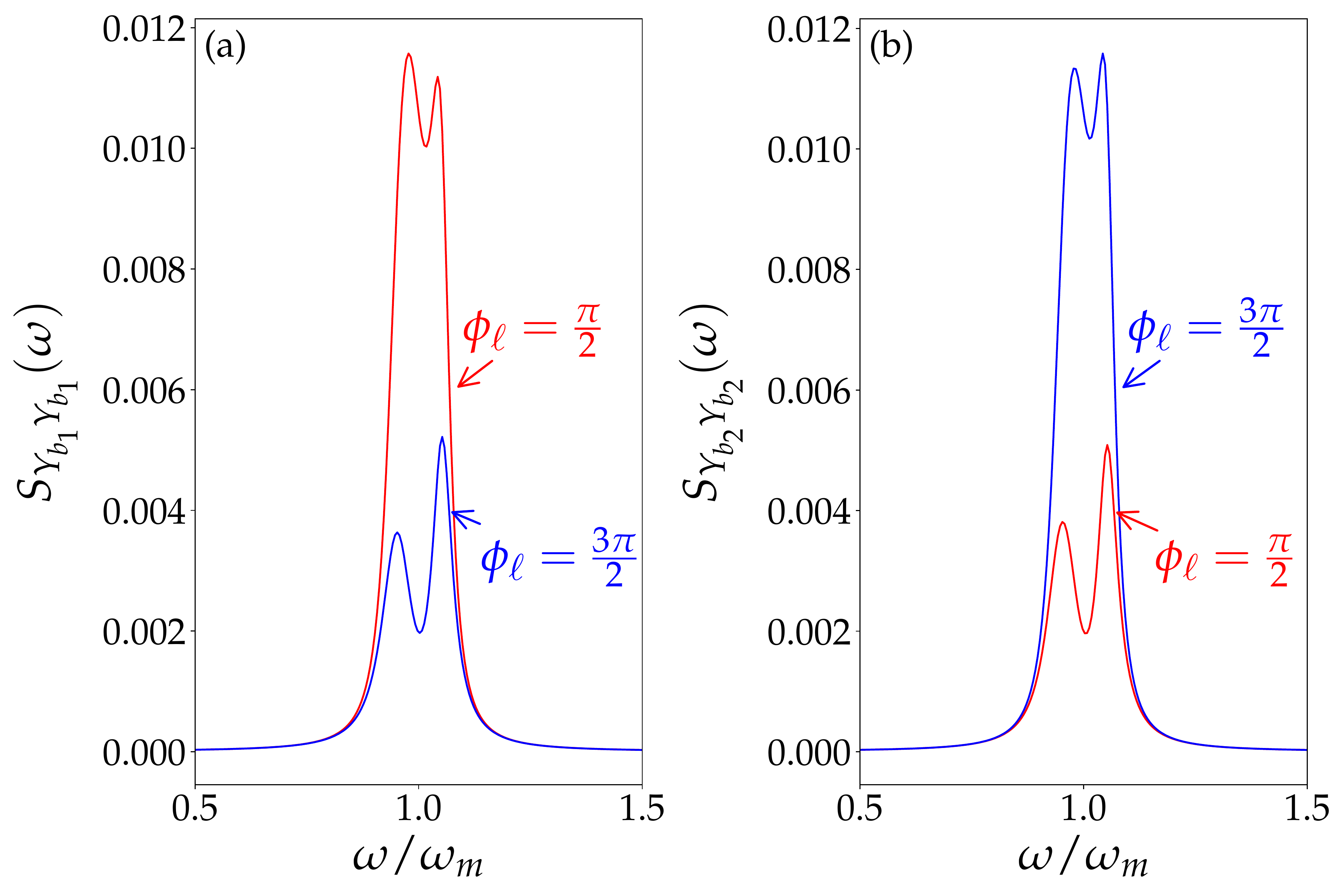}
      \caption{PSD of (a) first and (b) second mechanical resonators as a function of frequency for $\phi_{\ell}= \pi/2$ and $3\pi/2$. }
    \label{fig3}
\end{figure}

We initially focus on intracavity fluctuations and the backaction noise contribution from the second mechanical resonator to the cavity-output PSD. The first mechanical resonator’s contribution to the cavity-output PSD closely matches that of the second resonator, so it is not shown separately. This individual contribution is denoted as $S^{out}_{2 \rightarrow {cav}}(\omega) = |P_{Y_aY_{b_2}}(\omega)|^2 S^{in}_{Y_{b_2 b_2}} $. First, we begin by analyzing the intracavity spectrum for loop phases $\phi_{\ell} = 0,~\pi/2,~\pi$, and $3\pi/2$ [Fig.~\ref{fig2}(a)]. As a matter of fact, the spectrum for $\phi_{\ell}=3\pi/2$ is identical to that for $\pi/2$. Here, the main observation is that the intracavity noise spectrum can be switched from lower to upper band by tuning the loop phase from $\phi_{\ell}=0$ to $\pi$ while overall noise power essentially stays the same. This also applies to the cavity output as seen in Fig.~\ref{fig2}(b). The fact that the backaction noise spectrum can be swept in frequency by the loop phase is a ramification of backaction being a correlated noise \cite{PhysRevA.93.063809, Nielsen_2016, PhysRevLett.123.093602}. The backaction noise at $\phi_{\ell}=\pi/2,~3\pi/2$ displays the superpositions of $\phi_{\ell} = 0$ and $\pi$.
Before investigating the nonreciprocity in mechanical resonators, we first examine the internal noise PSD in Fig.~\ref{fig3}. It is evident that by appropriately selecting the loop phase as $\phi_{\ell} = \pi/2$ and $3\pi/2$, one can boost the internal noise of the first and second mechanical resonator, respectively. This corroborates our earlier report that, at these loop phases,
one of the mechanical resonators can be selectively cooled,\cite{sutluouglu2024selective}
highlighting the role of the synthetic phase in regulating fluctuations.

\begin{figure}[H]
  \centering
  \includegraphics[width=\linewidth]{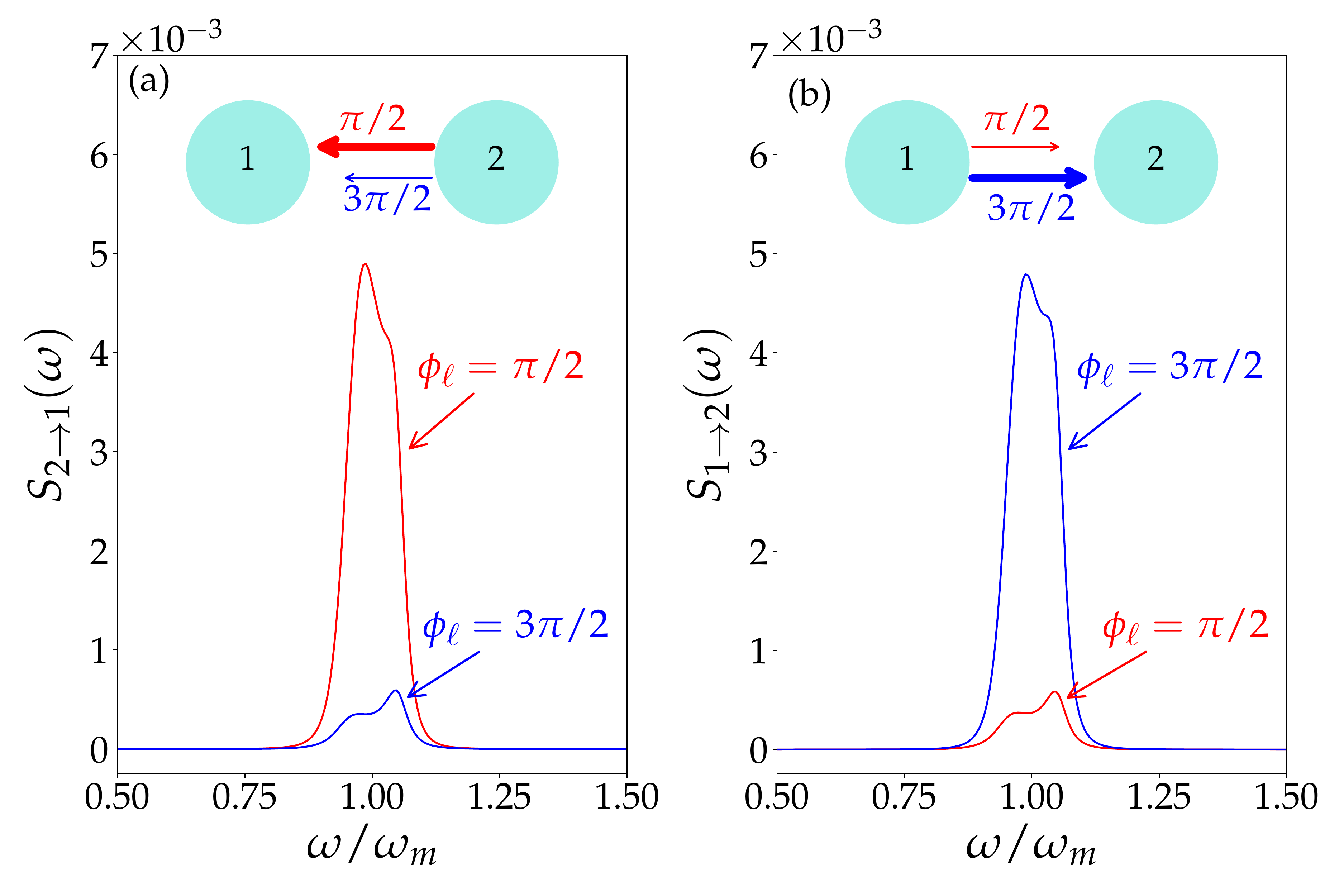}
      \caption{Noise transfer pathways between the two mechanical resonators.
    (a) Contribution of the second resonator’s fluctuations to the momentum-noise spectrum of the first resonator,
    (b) contribution of the first resonator’s fluctuations to the momentum-noise spectrum of the second resonator for $\phi_{\ell} = \pi/2$ and $3\pi/2$. The arrows in the insets depict the weight of noise transfer between mechanical resonators for $\phi_{\ell} = \pi/2$ and $3\pi/2$.    }
    \label{fig4}
\end{figure}

\begin{figure}[H]
  \centering
  \includegraphics[width=\linewidth]{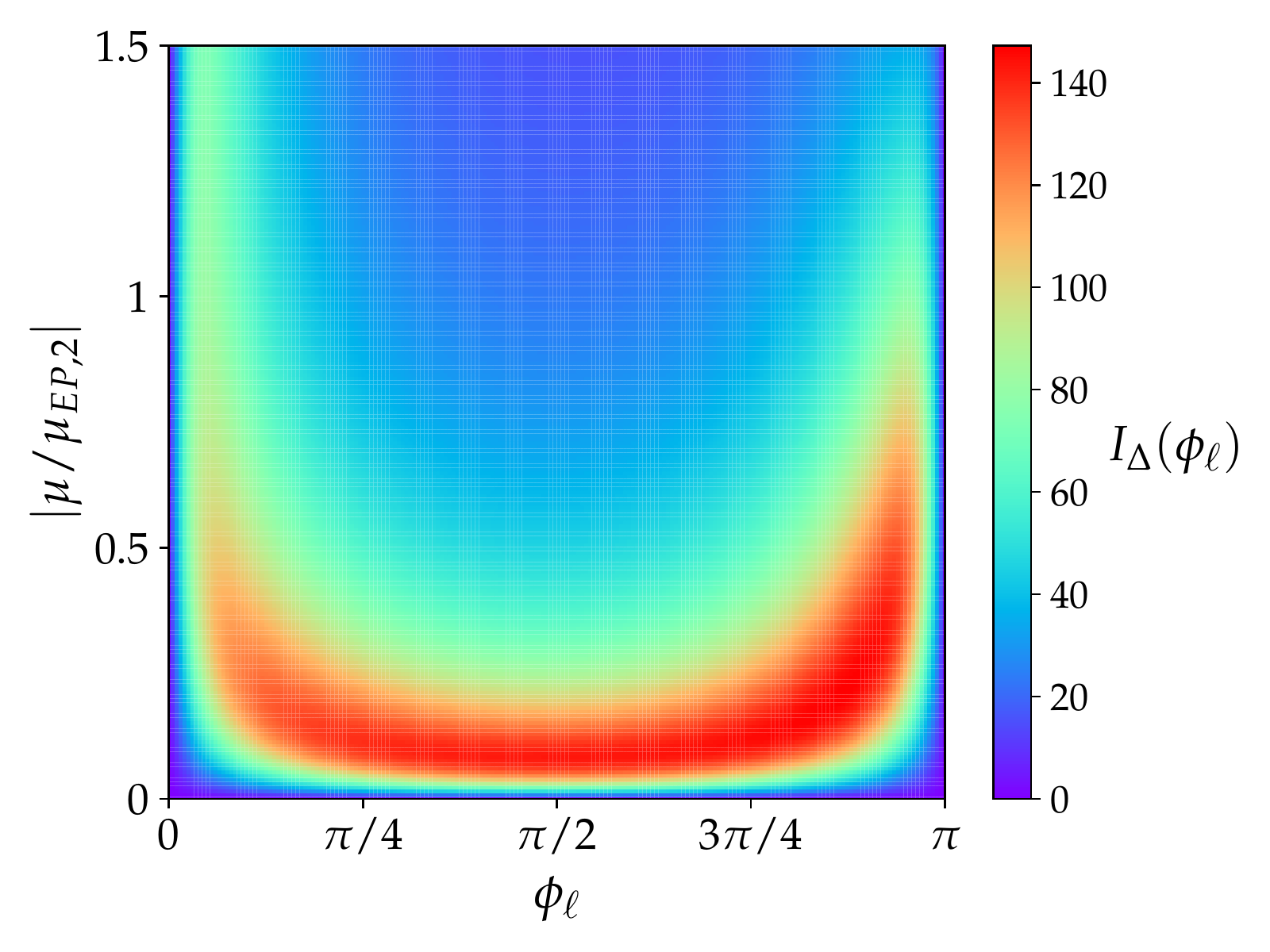}
      \caption{Nonreciprocity measure $I_\Delta(\phi_{\ell})$ as a function of the loop phase $\phi_{\ell}$ (horizontal axis) and mechanical coupling constant normalized to second EP $|\mu/\mu_{EP,2}|$ (vertical axis). Dark-red regions indicate stronger nonreciprocity. }
    \label{fig5}
\end{figure}
\begin{figure}[H]
  \centering
  \includegraphics[width=\linewidth]{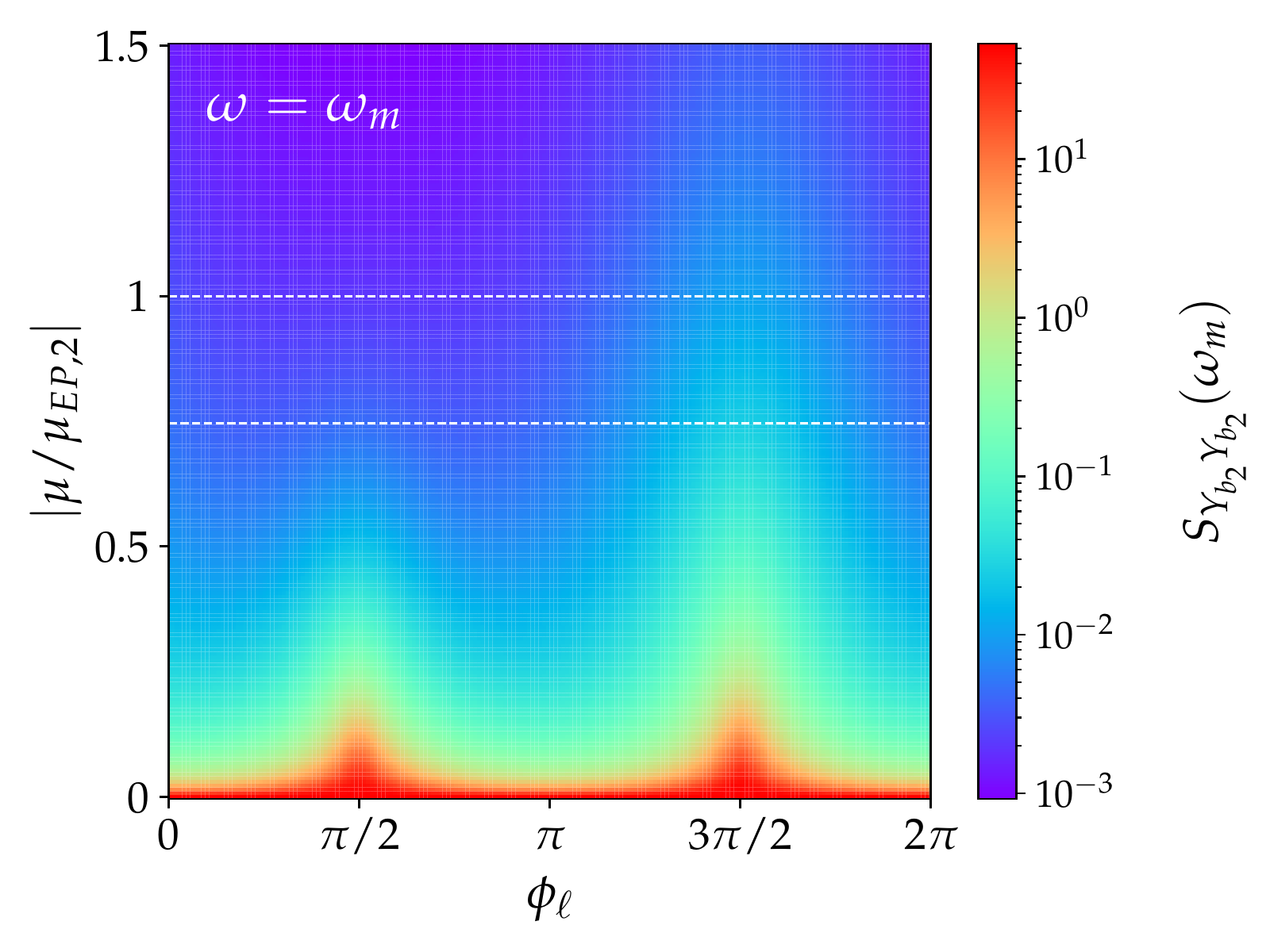}
      \caption{PSD of second mechanical resonator with respect to loop phase $\phi_{\ell}$ (horizontal axis) and mechanical coupling constant normalized to second EP $|\mu/\mu_{\mathrm{EP},2}|$ (vertical axis) at $\omega = \omega_m$. Horizontal white-dashed lines mark the first and second EPs, $|\mu_{\mathrm{EP},1}|$ and $|\mu_{\mathrm{EP},2}|$. Dark-red regions correspond to areas of enhanced noise in the PSD.}
    \label{fig6}
\end{figure}
\begin{figure}[H]
  \centering
  \includegraphics[width=\linewidth]{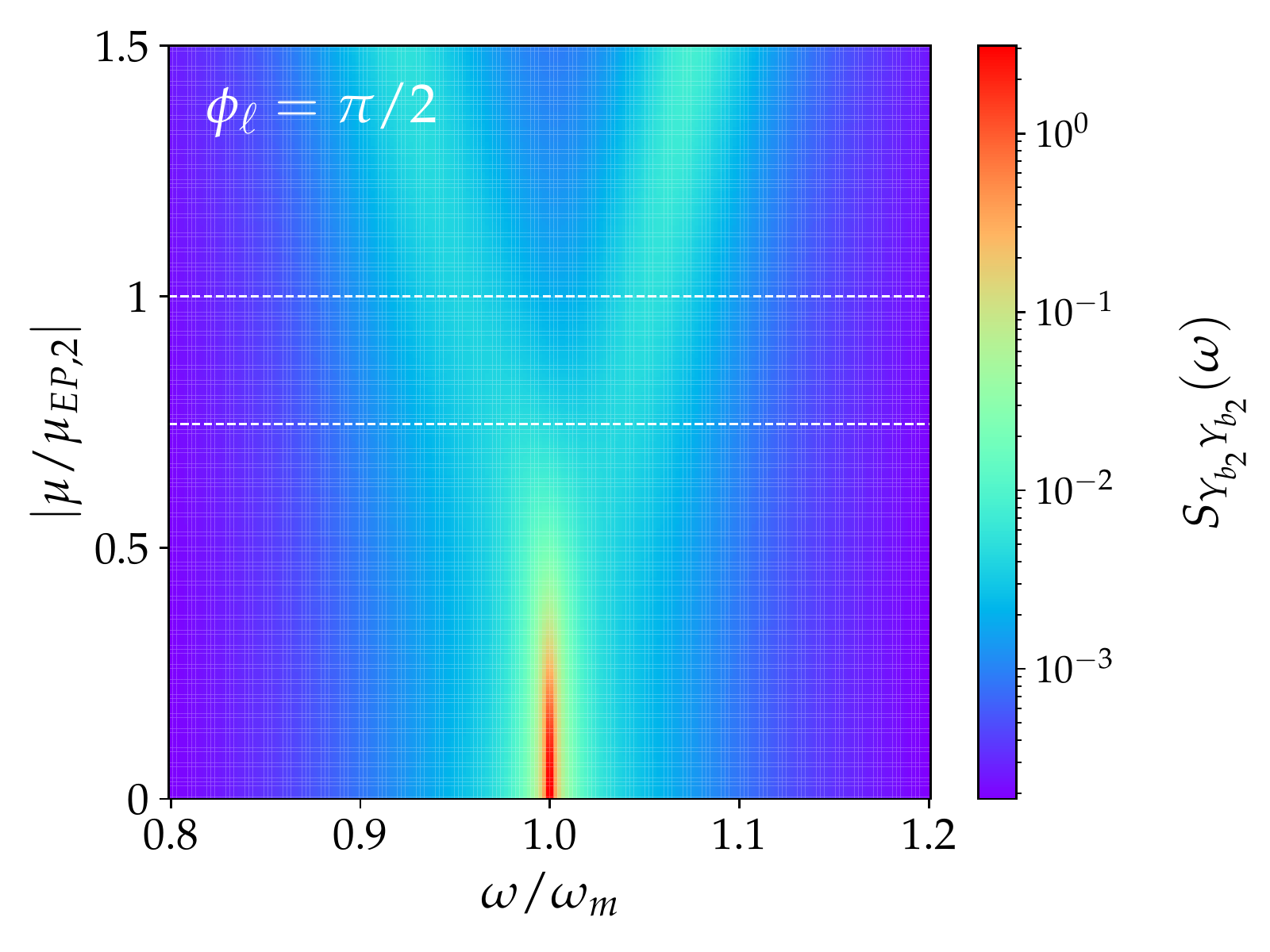}
      \caption{PSD of second mechanical resonator as a function of frequency (normalized to $\omega_m$, horizontal axis) and mechanical coupling constant normalized to second EP $|\mu/\mu_{\mathrm{EP},2}|$ (vertical axis) for $\phi_{\ell}= \pi/2 $. Horizontal white-dashed lines mark the first and second EPs, $|\mu_{\mathrm{EP},1}|$ and $|\mu_{\mathrm{EP},2}|$. Dark-red regions correspond to enhanced noise levels, while dark-blue regions indicate reduced noise. }
    \label{fig7}
\end{figure}

In Fig.~\ref{fig4}, we plot the individual noise transfers for loop phases, $\phi_{\ell}= \pi/2$ and $3\pi/2 $: (a) from the second mechanical resonator to the first $S_{2 \rightarrow 1}(\omega) = |T^\prime_{Y_{b_1}Y_{b_2}}(\omega)|^2 S^{\mathrm{in}}_{Y_{b_2}Y_{b_2}}$, and (b) from the first mechanical resonator to the second $S_{1 \rightarrow 2}(\omega) = |T^\prime_{Y_{b_2}Y_{b_1}}(\omega)|^2 S^{\mathrm{in}}_{Y_{b_1}Y_{b_1}}$, respectively. Here, we see that there is a nonreciprocal noise transfer depending on the loop phase. Whenever this phase is $\pi/2$, noise flow from the second to the first mechanical resonator is dominant, and vice versa for $3\pi/2$. This result is also consistent with earlier finding where we achieved a dominant ground state cooling of second and first mechanical resonators when loop phase is chosen as $\phi_{\ell}=\pi/2$ and $\phi_{\ell}=3\pi/2$, respectively \cite{sutluouglu2024selective}. These figures illustrate the system’s behavior at the second EP, $\mu_{\mathrm{EP},2}$; we refer to Appendix~A for additional plots at different mechanical coupling constants. Subsequently, we integrate these spectral contributions to obtain the nonreciprocal noise measure. Hence, we investigate $I_\Delta(\phi_{\ell})$ as a function of the mechanical coupling constant normalized to $\mu_{\mathrm{EP},2}$ and the loop phase in Fig.~\ref{fig5}. Naively, one would expect a characteristic that is symmetric about $\pi/2$; however, nonreciprocity measure exhibits a pronounced asymmetry, which is caused by the non-trivial loop phase dependence of three-mode coupling; further details are provided in Appendix~B. $I_\Delta(\phi_{\ell})$ increases sharply as soon as the loop phase $\phi_{\ell}$ is detuned even slightly from $0$ or $\pi$, reflecting the key role of the broken time-reversal symmetry. As the mechanical coupling constant is increased, the system gradually regains reciprocity, and the nonreciprocal response diminishes. However, their significance becomes evident in Figs.~(\ref{fig6}) and (\ref{fig7}) where we plot the momentum quadrature PSD of the second mechanical resonator for analyzing the internal fluctuations.

Analyzing the PSD of internal fluctuations is complicated due to their intricate dependence on multiple parameters such as $\omega$, $\mu$, and $\phi_{\ell}$. Therefore, Figs.~\ref{fig6} and \ref{fig7} are to be considered as two complementary figures, first one is at $\omega = \omega_m$ as a function of $\phi_{\ell}$ at fixed $\mu$, and the second one is at $\phi_{\ell}=\pi/2$ as a function $\omega$ at fixed $\phi_{\ell}$. The behavior of PSD with varying loop phase in Fig.~\ref{fig6} agrees with Figs.~(\ref{fig3}) and (\ref{fig4}), which confirms that noise in mechanical resonators can be controlled with loop phase. According to Fig.~\ref{fig5}, the nonreciprocity measure decreases when the system approaches EPs, whereas Fig.~\ref {fig7} suggests that sensitivity increases with the increasing $\mu$. Therefore, devices like isolators require a smaller mechanical coupling constant for a nonreciprocal noise flow, but to achieve a better sensitivity (reduced noise PSD at $\omega_m$), a system operating closer to the EP will be much more convenient for sensors. Fig.~\ref{fig7} shows that  PSD develops a characteristic double-peak structure as $\mu$ approaches $\mu_{EP,1}$, similar to our earlier study where a transition from single to double transparency peak was observed at the EP \cite{PhysRevA.104.033504}. For smaller values of $\mu$, noise becomes dominant at the resonance frequency $\omega_m$. 

\section{Conclusions}
In summary, we have conducted a theoretical investigation of a three-mode closed-loop optomechanical system that exhibits a synthetic magnetic field. While the Onsager-Casimir reciprocity relations remain valid, the introduction of a loop phase effectively breaks time-reversal symmetry, thereby enabling nonreciprocal propagation of backaction noise within the system. By analyzing the PSD of the internal fluctuations, we demonstrated that this nonreciprocity is manifested as a directional transfer of noise between the mechanical resonators. This directional behavior is a direct consequence of the phase-dependent interference, which redistributes correlated noise asymmetrically across the system's resonators. An equally vital point is that the experimental realization of this theoretical proposal is currently feasible. For instance, the loop phase can be imparted as a phase difference between the two mechanical resonators by a pair of vibrating dielectric membranes embedded in a Fabry-Pérot cavity, with their individual radiative coupling amplitudes and phases tunable through the membranes’ absolute and relative positions \cite{PhysRevA.99.023851, piergentili2018two}. Furthermore, our scheme demands a strong intermechanical coupling constant, which has been demonstrated in earlier experiments as reviewed in the preceding experimental-realization Sec.~\ref{sec3}.

Importantly, this nonreciprocal noise transfer can be exploited for practical applications, such as enhancing the performance of optomechanical sensors. In particular, increasing the intermechanical coupling constant leads to improved sensitivity by amplifying the noise contrast between the resonators. Conversely, if the primary goal is to maximize nonreciprocity rather than sensitivity, a smaller intermechanical coupling becomes advantageous, as it enhances the asymmetry in the noise flow. These findings highlight the tunability of nonreciprocal effects in such systems and offer valuable insights for the design of next-generation quantum sensors and signal-processing devices. 

\begin{acknowledgments}
We thank İnanç Adagideli for fruitful discussions. B.S.E. and C.B. acknowledge support from Air Force Office of Scientific Research (AFOSR) Grant No. FA9550-22-1-0444.
S.K.O. acknowledges support from Air Force Office of Scientific Research (AFOSR) Multidisciplinary University Research Initiative (MURI) Award No. FA9550-21-1-0202.
\end{acknowledgments}

\appendix
\section*{Appendix A: Results under different mechanical coupling constants}
\label{appendix:a}
\begin{figure}[H]
  \centering
  \includegraphics[width=\linewidth]{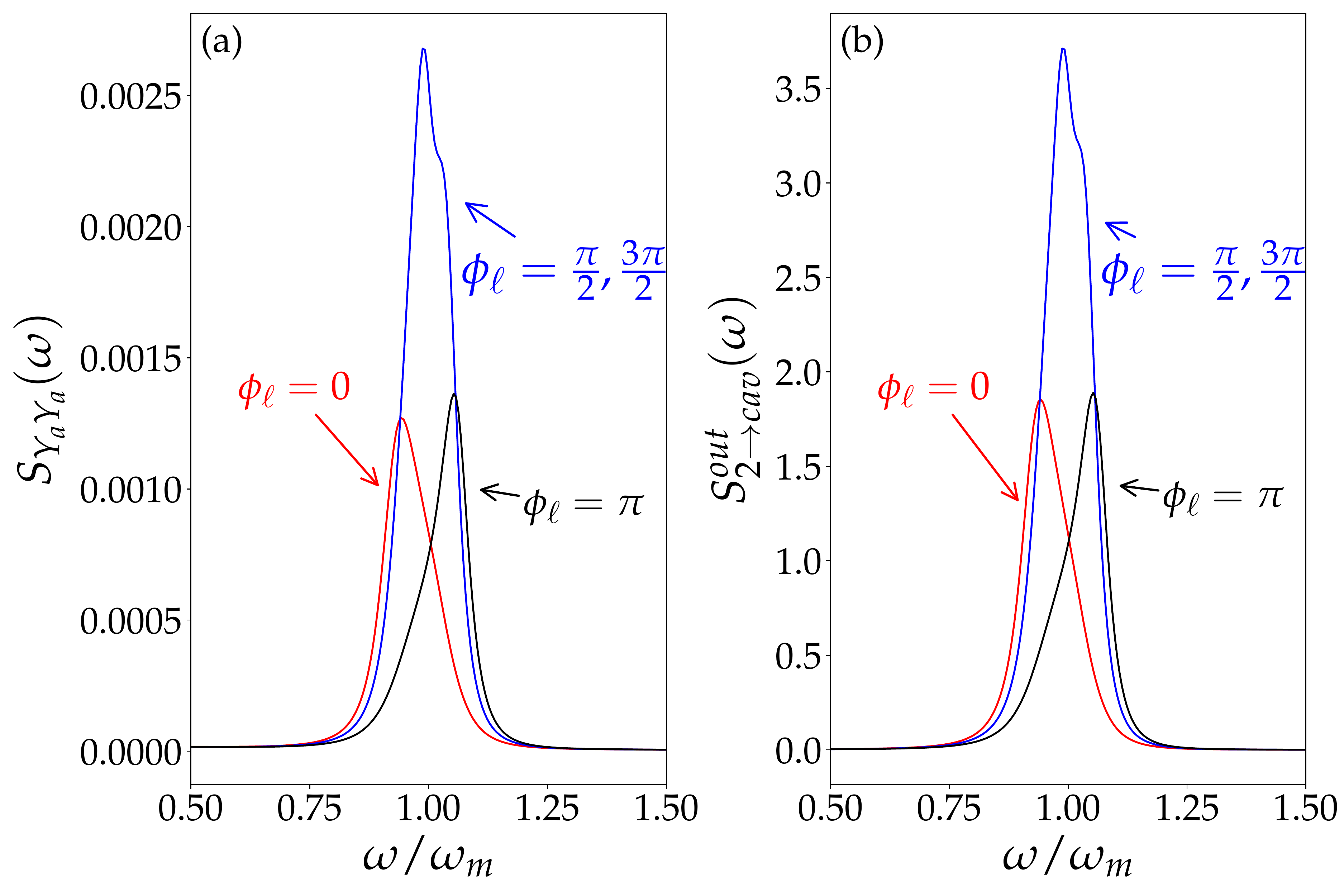}
      \caption{(a) Optical intracavity PSD, (b) individual noise contribution of the second mechanical resonator to the cavity-output noise spectrum for three different loop phases, $\phi_{\ell} = 0,~\pi/2,~\pi$, and $3\pi/2$ at $\mu = \mu_{EP,1}$.}
    \label{app1}
\end{figure}

\begin{figure}[H]
  \centering
  \includegraphics[width=\linewidth]{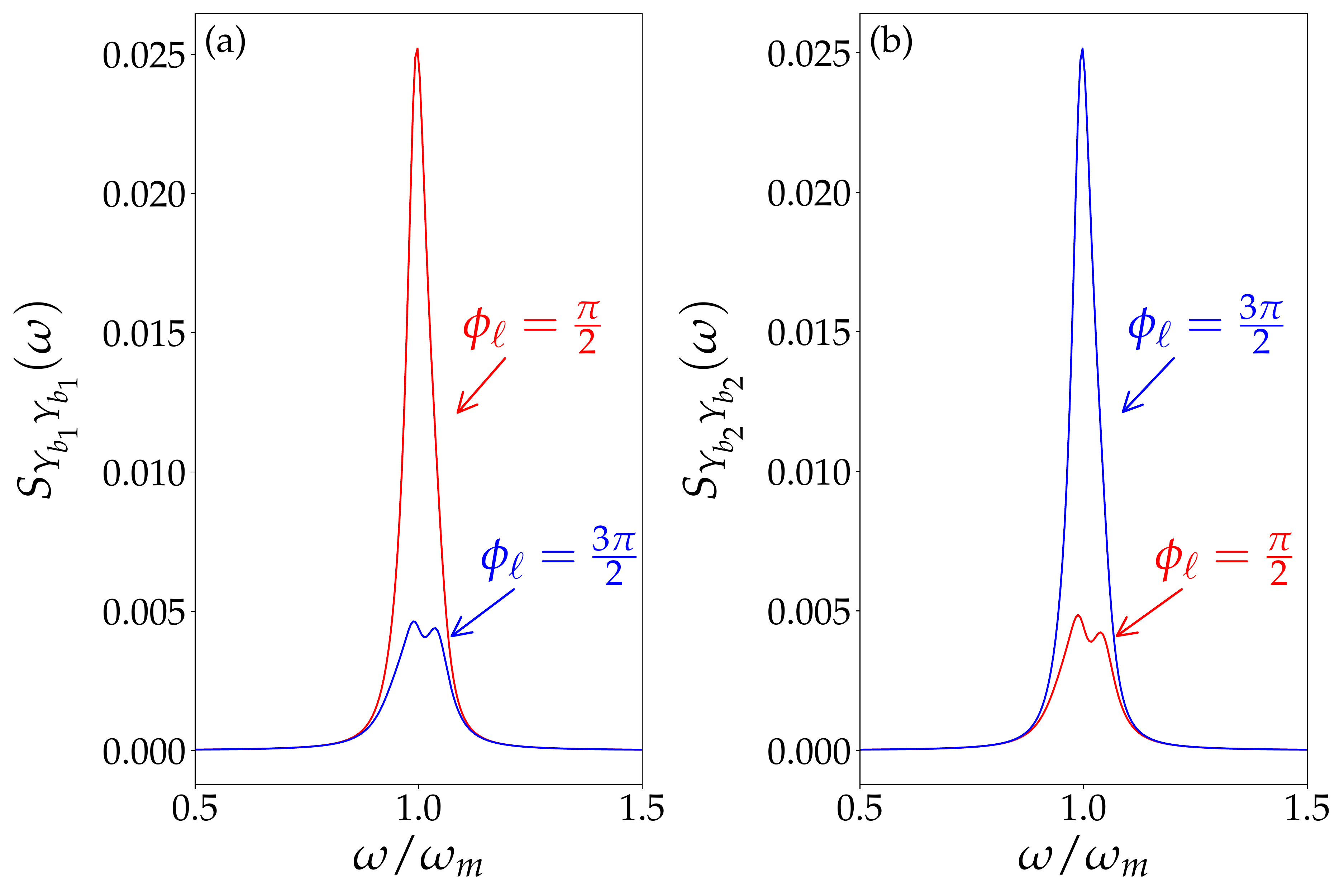}
      \caption{PSD of (a) first and (b) second mechanical resonators as a function of frequency for $\phi_{\ell}= \pi/2$ and $3\pi/2$ at $\mu = \mu_{EP,1}$.}
    \label{app2}
\end{figure}

\begin{figure}[H]
  \centering
  \includegraphics[width=\linewidth]{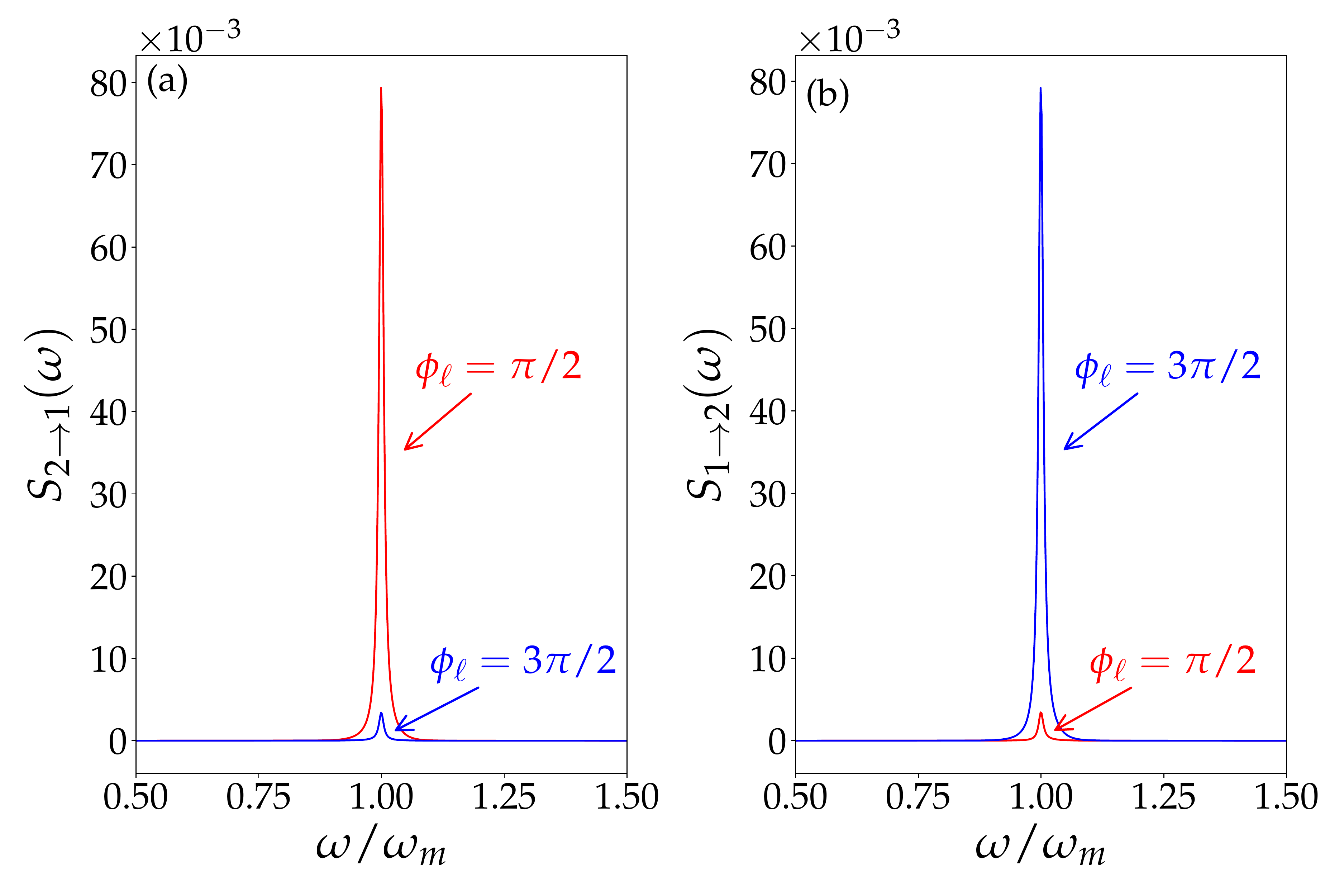}
      \caption{Noise transfer pathways between the two mechanical resonators.
    (a) Contribution of the second resonator’s fluctuations to the momentum-noise spectrum of the first resonator,
    (b) contribution of the first resonator’s fluctuations to the momentum-noise spectrum of the second resonator for $\phi_{\ell} = \pi/2$ and $3\pi/2$ at the $\mu = 0.5\mu_{EP,1}$.}
    \label{app3}
\end{figure}

\begin{figure}[H]
  \centering
  \includegraphics[width=\linewidth]{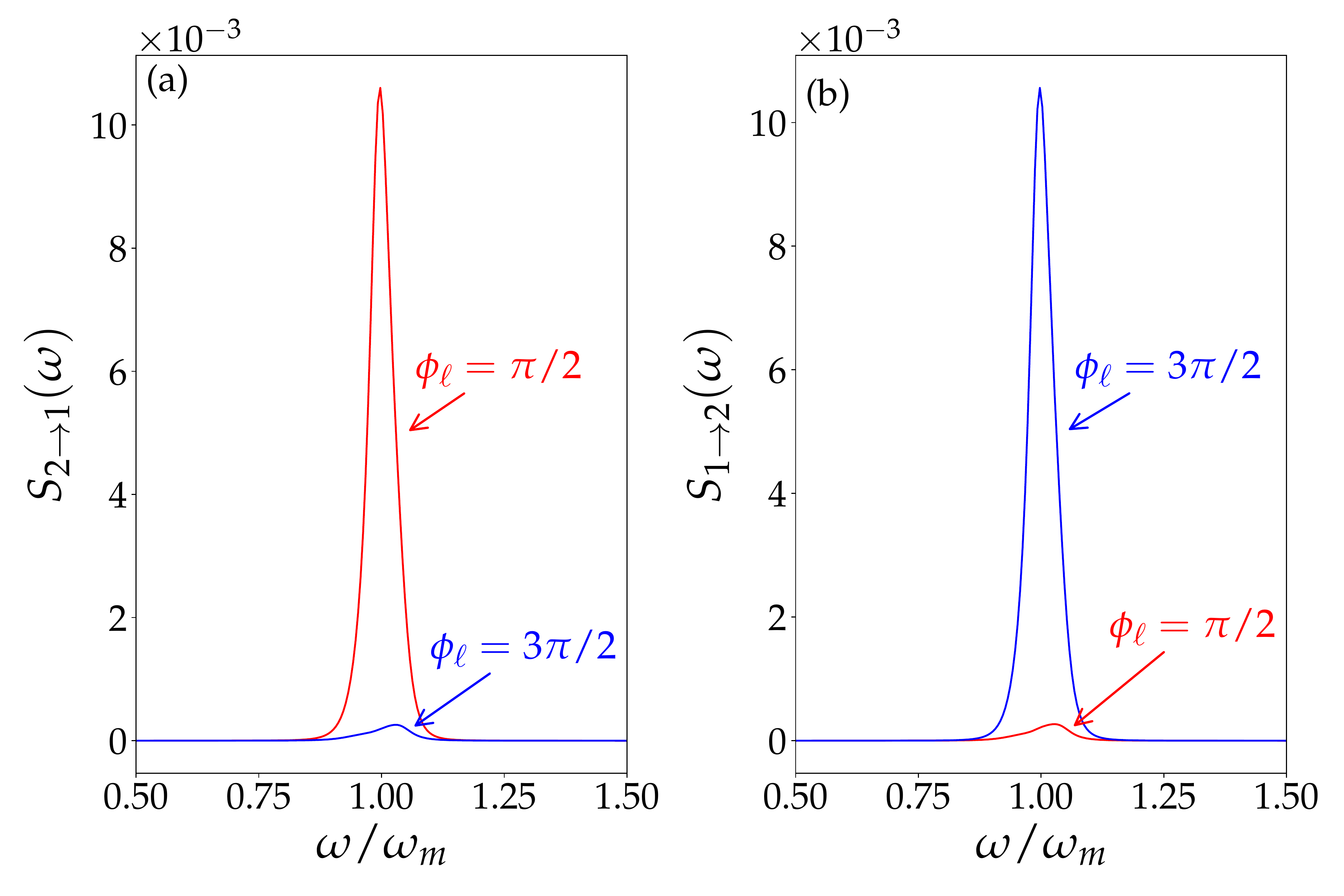}
      \caption{Noise transfer pathways between the two mechanical resonators $\phi_{\ell} = \pi/2$ and $3\pi/2$ at $\mu = \mu_{EP,1}$.}
    \label{app4}
\end{figure}

\begin{figure}[H]
  \centering
  \includegraphics[width=\linewidth]{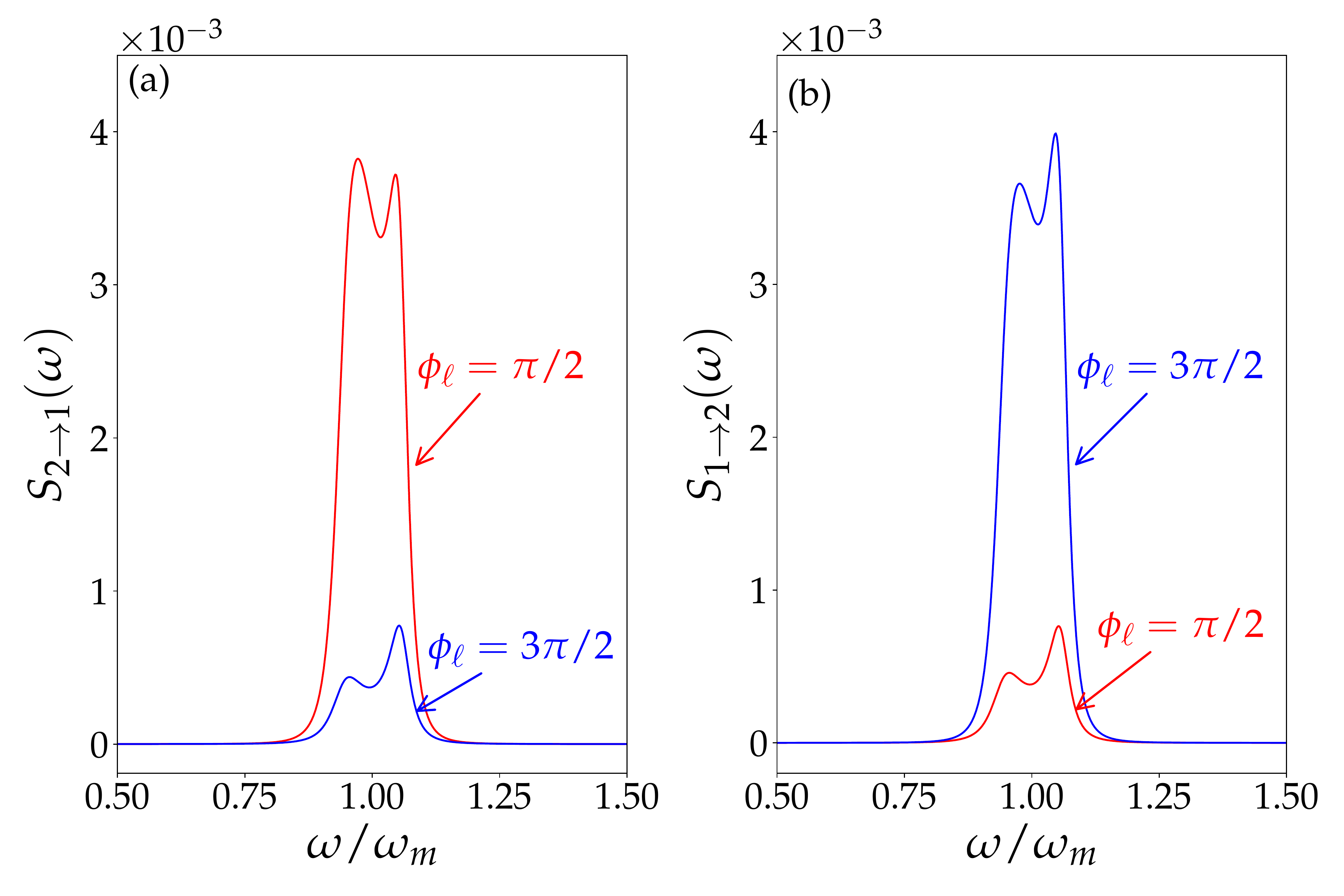}
      \caption{Noise transfer pathways between the two mechanical resonators $\phi_{\ell} = \pi/2$ and $3\pi/2$ at $\mu = 1.5 \mu_{EP,1}$.}
    \label{app5}
\end{figure}

\begin{figure}[H]
  \centering
  \includegraphics[width=\linewidth]{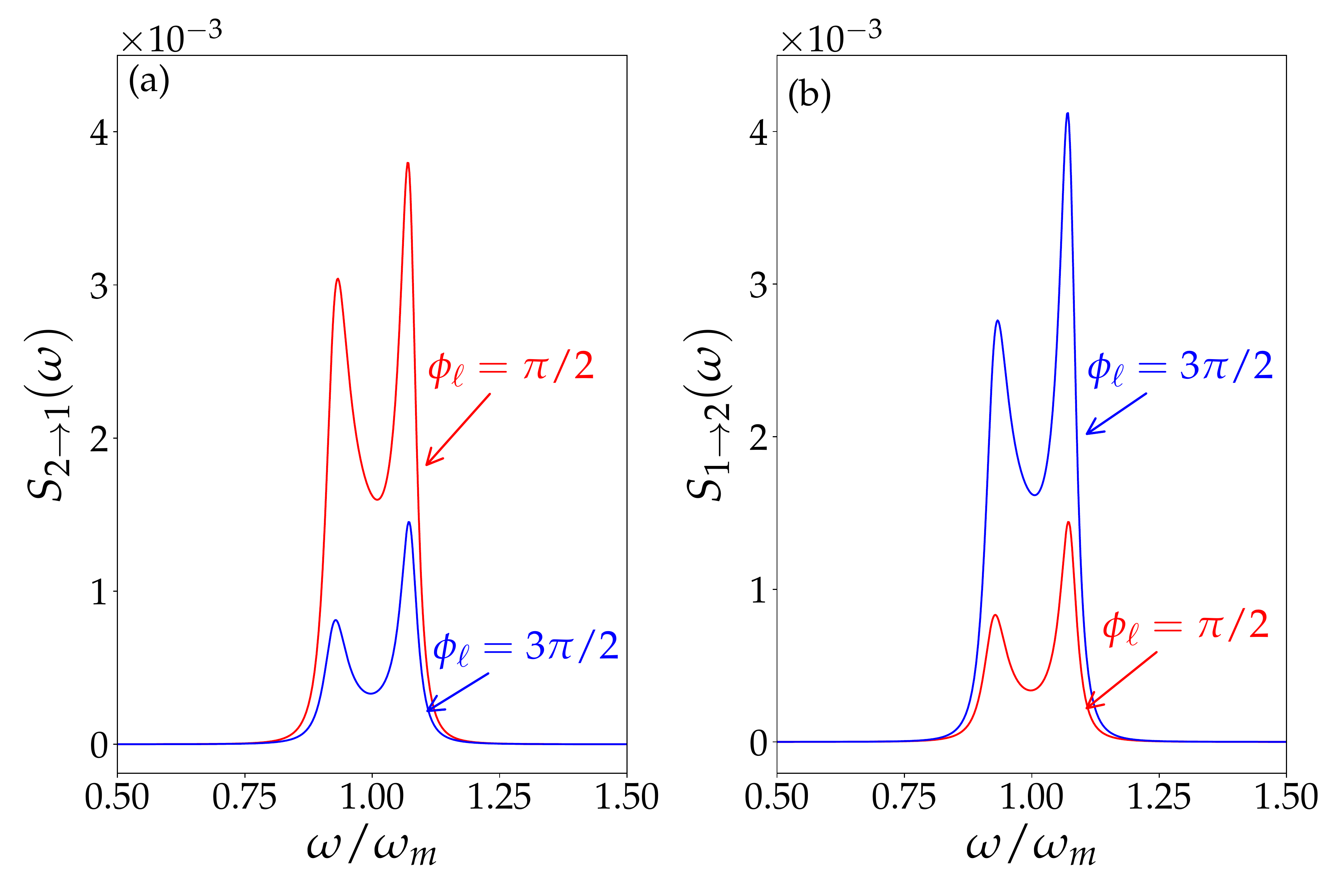}
      \caption{Noise transfer pathways between the two mechanical resonators $\phi_{\ell} = \pi/2$ and $3\pi/2$ at the $\mu = 1.5\mu_{EP,2}$.}
    \label{app6}
\end{figure}

\begin{figure}[H]
  \centering
  \includegraphics[width=\linewidth]{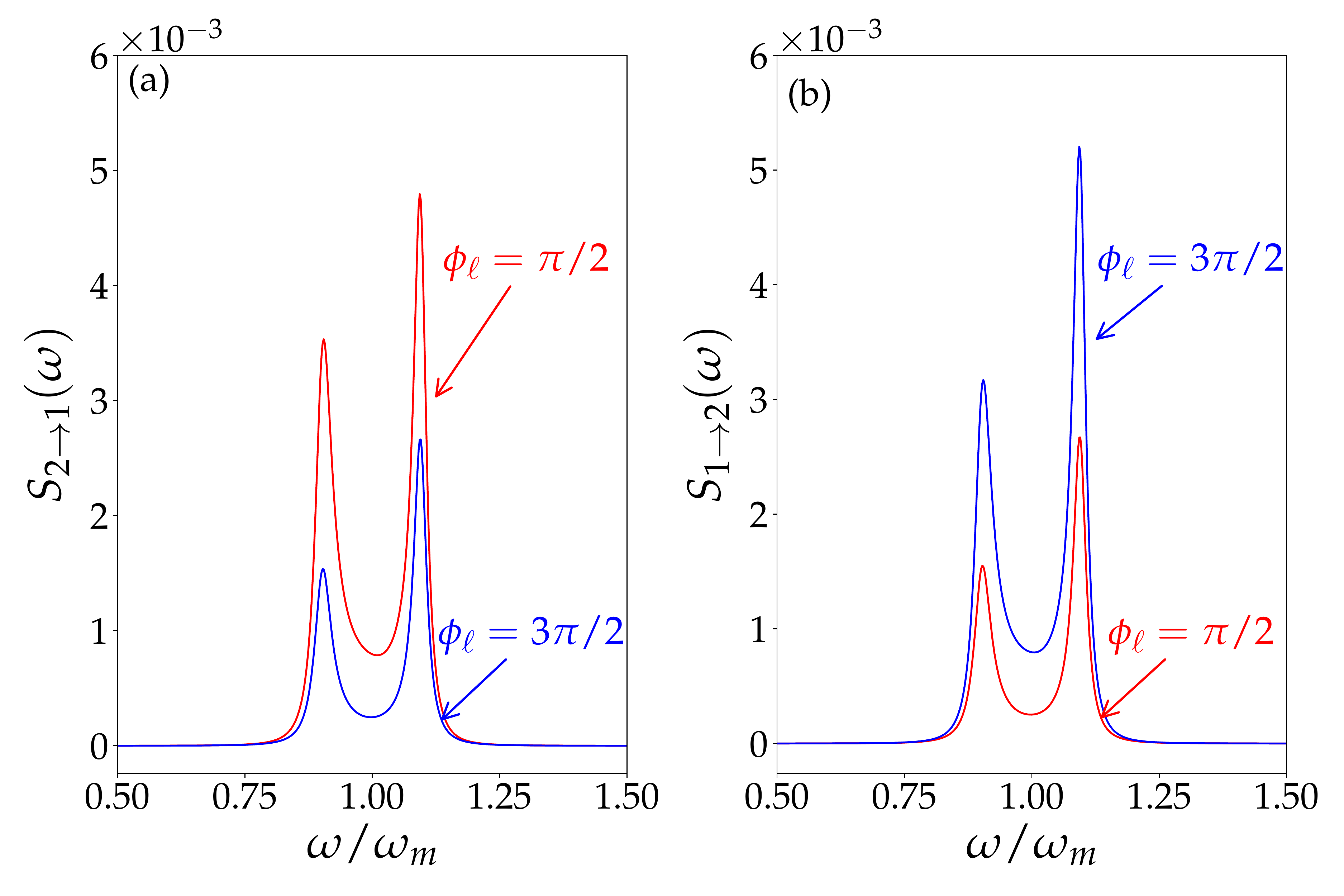}
      \caption{Noise transfer pathways between the two mechanical resonators $\phi_{\ell} = \pi/2$ and $3\pi/2$ at the $\mu = 2\mu_{EP,2}$.}
    \label{app7}
\end{figure}

In this Appendix, we elaborate on the effects of proximity to either of the EPs by varying the intermechanical coupling constant. For this purpose, we replot some of the figures in the Results section for different $\mu$ values. In Fig.~\ref{app1}, we plot intracavity fluctuations and the noise contribution of the second mechanical resonator to cavity-output PSD at $\mu_{EP,1}$ to compare with Fig.~\ref{fig2}. In comparison, the double peak characteristic in PSD when $\phi_\ell = \pi/2, 3\pi/2$ is not fully developed at $\mu_{EP,1}$, whereas the noise power increases slightly. 
In Fig.~\ref{app2}, we examine internal mechanical fluctuations at the $\mu_{EP,1}$ compared to results at the $\mu_{EP,2}$. This outcome aligns well with the findings in Fig.~\ref{fig5}, which shows enhanced sensitivity for higher values of the mechanical coupling constant. 
In Figs.~\ref{app3}-\ref{app7}, we replot Fig.~\ref{fig4} to analyze the differences depending on the proximity of EPs. These figures serve to visually support the discussion presented in Figs.~\ref{fig6} and \ref{fig7}. The nonreciprocity measure given in Eq.(\ref{non-reciprocity measure}) is defined as the difference between the integrated PSD. Hence, it increases for small values of $\mu$ since noise power increases. However, for larger mechanical coupling constant values, noise power decreases, and as a result, integrated noise power also decreases. Another notable feature is the emergence of a double peak in the PSD as the system approaches the EPs, which can also be inferred from Fig.~\ref {fig7}. 

\section*{Appendix B: Root Loci of the upper half-plane eigenvalues}
\label{appendix:b}
To identify the origin of asymmetry with respect to $\phi_{\ell}=\pi/2$ in Fig.~\ref{fig5}, in this appendix we plot real and imaginary parts of the upper half-plane eigenvalues $(z(|\mu|,\phi_{\ell}) = \alpha +i\omega,~\omega>0) $ of the drift matrix $\mathbf{M}$ as the loop phase varies $\phi_{\ell} \in [0,\pi]$ (for $\phi_\ell \in [\pi, 2\pi]$ the same loci are retraced backwards). These are displayed at two different intermechanical coupling values chosen to be at either of the two EPs. For any $\phi_{\ell}$ there exist three eigenvalues corresponding to supermodes formed by the hybridization of two mechanical and one photonic modes. The conspicuous asymmetric behavior of these supermodes with respect to $\phi_{\ell}=\pi/2$ (marked with vertical dashed lines) in Fig.\ref{app8} is the source of lack of this symmetry in the nonreciprocity measure in Fig.~\ref{fig5}. 
\begin{figure}[H]
  \centering
  \includegraphics[width=\linewidth]{./Figures/Fig15.pdf}
      \caption{Real and imaginary parts of complex upper half plane eigenvalues $z(|\mu|,\phi_{\ell})$ of the drift matrix $\mathbf{M}$ as loop phase $\phi_{\ell}$ varies from $0$ to $\pi$ at the EPs, $|\mu_{EP,1}|$ and $|\mu_{EP,2}|$. A vertical axis is placed at $\phi_{\ell}=\pi/2$ to highlight the asymmetry. }
    \label{app8}
\end{figure}

\end{document}